# Reflections on digital innovation

Eric Monteiro, Norwegian University of Science and Technology (NTNU),
https://www.ntnu.no/ansatte/ericm

**Abstract**. The paper by Henfridsson et al. opens up a new agenda for IS research on the content and process of digital innovation. The crucial element in their perspective is the role of recombination in innovation. They supplement an emphasis on design recombination with a symmetrical emphasis on use recombination. While supporting Henfridsson et al.'s overall argument, I point out how central parts overlap with and are extended in disciplines outside IS research.

## Introduction

The paper lays out a generative research programme for understanding our engagement with digital technology (resources). Driven by an urge to increase our analytic precision in describing the phenomenon, the paper offers a helpful roadmap to empirically analyse the dynamics of digital innovation. The paper is a much-needed rethinking of our engagement with digital technologies and the way innovation in/ with digital technologies unfolds. The paper's ambition is to lay the groundwork for subsequent IS research. The paper provides a framework of the ingredients, actors, processes and outcomes of digital innovation.

There is a lot to like about the paper. My reflections obviously mirror my background. My commitment with IS research is characterized by an extensive engagement with ideas from other disciplines, notably science and technology studies (STS), innovation studies and cultural theory. In an effort to highlight what I take to be the key parts of the paper's core idea – an account of value generation and consumption in digital innovation – I explain how the authors resonate with ideas from disciplines outside of IS research. I take this affinity with other disciplines to represent a productive enrichment of the dominant IS based accounts. Much of the vibrancy and energy in IS research, it seems to me, relies on the open economy of ideas IS research enjoys with other disciplines.

In the following, I focus on three salient ideas of the paper and critically challenge these by drawing on the discourses in non-IS disciplines, primarily STS, from where they originate.

## "IS research have over-emphasized design over use"

*The idea*: The paper points out how dominant perspectives in IS research have privileged design over use of technology. Despite nods to the contrary, the bias persists. It is, the paper notes, more often implicit than explicit. The paper argues for a design-over-use bias by identifying it in the least likely place in line with a logic of critical case (Flyvbjerg 1998) Also in the value cocreation and service innovation literature, the authors argue, there is a design bias through the emphasis on 'resource integration' performed by firms rather than users:

> "this firm-centric, design emphasis also pervades IS research on cocreation (Grover and Kohli 2012; Sarker et al. 2012), where recombination, consistent with a service- dominant view (Lusch and Nambisan 2015), is understood as 'resource integration' …[where] firms [through design] integrate resources to create an attractive offering" (pp. 5-6)

*The link*: Readers familiar with STS would have no quarrel with the paper's critique of over-emphasizing design. In fact, the very motivation and background of STS could be seen to rely on this critique. It was the cornerstone of STS' debunking of technological determinism in all its manifestations. In STS, the critique of the over-emphasis on design was part of the political project of demonstrating that design was not the exclusive task of privileged designers imposing technology on docile users. Design as conceptualized in STS leaves a significant amount of leeway to the users' creativity and ingenuity. In STS, the proactive engagement of users is a necessary part of innovation. The conceptual dichotomy between design and use is dismantled. The core of the STS programme amounts to demonstrating 'use recombination' empirically. The example of the Zimbabwean bush pump is illustrative (de Laet and Mol, 2000). It is a water pump in use in several dozen sub-Saharan countries in Africa. The water pump is assembled in any number of configurations by extensive use recombination in the local communities. Use recombination are so radical that the Zimbabwe push pump should be seen as 'fluid' similar to the characterization in the paper.

*The issue*: We all agree users' influence is vital. The principle issue with the critique of over-emphasis on design, subject to heated debates in STS but not addressed in the paper, is whether there are any *limits* to the influence of users. This raises the concern of social determinism i.e. that 'all' is in the hands of users. The paper could at times be seen to wander dangerously close to such a position, notably when explaining the 'agnostic' nature of digital technology:

> "Digital resources are agnostic in the sense that their meaning in the use situation is largely defined by their relationships to other resources" (p. 3) and that "any digital resource…may become a constituent of many different user value paths as it is recombined and made meaningful by different users" (p. 5).

In escaping the design bias of digital innovation found in IS research, the challenge for the paper is to avoid falling into the opposite ditch i.e. social determinism. Users have extensive, not unbridled, leeway for manoeuvring.

**"The term 'use' of digital technology is crude"**

*The idea*: The second idea in the paper is that it dismisses the term 'use' of digital technology as overly crude. There is a need, the paper notes, to "increase the granularity" of how we conceptualise 'use'. IS research, targeting the "rich, emergent interactions" between technology and organizations as made clear in Lee's (1999, p. v) inaugural MISQ editorial, has a crude conceptualization of one of its central constructs, 'use' of technology. To this end, the paper decomposes a monolithic notion of the 'system' (or the 'technology') into its constituent parts, what the paper coins digital resources. As succinctly put in the paper, "'use' is no longer a discrete act" (p. 8) i.e. it is no longer a monolithic concept.

*The link*: To merely state that a group 'use' the system/ technology hides as much of the empirical phenomenon as it reveals. It fails to differentiate between the empirically distinct situations depending on *what* parts of the system was used, *whom* in the group used what, for what *purpose*, *frequency* of use and *evolution* of use patterns over time. In recognition of this, Leonardi (2012) analyses how distinct functions (corresponding to digital resources) are used over time across different groups (for others who have underscored the need to differentiate more fine-grained, see Barley 1996 or Monteiro and Hanseth 1996).

*The issue*: A critique against the crudeness of the notion of 'use' is clearly a necessary and important prerequisite for encouraging a more nuanced IS research appreciation of our engagement with digital technologies (resources). I whole-heartedly support it. My concern is whether this critique in its present formulation in the paper is enough. In cultural studies, this critiqued is pushed further by challenging residues of functionalistic reasoning. 'Use' of technology, cultural studies argue, need to take on board also symbolic aspects. For instance, Silverstone and Haddon (1996) proposes the notion of domestication of technology. Domestication is the process over time through which technology is woven into the moral economy of our everyday lives. We do not 'use' technology, we *live* with technology (cf. Faraj et al. 2016).

**"There is a multiplicity of values of technology"**

*The idea*: The launching pad for the paper is the stipulation that neither design nor use is monolithic but rather are recombination of the constituent parts (i.e. digital resources). Given this, the third idea of the paper is the emphasis of the multiplicity of possibilities of recombination for both design and use as "[i]f a digital resource is part of multiple value paths, it can then assume different functions depending on the way it relates to other resources" (p. 4) and any digital resource "may become a constituent of many different user value paths as it is recombined and made meaningful by different users" (p. 5). This entails that "a specific digital resource may be part of several different value paths" (p. 11) and that "the meaning, or function, of a specific digital resource changes in tandem with its relations to the other resources of a value path" (p. 11).

*The link*: The paper's understanding of digital innovation is strongly influenced by perspectives on digital *platforms*. The paper's idea of a multiplicity of values may be understood as a critique of, to use Law's (1999) phrase, the 'centred' bias of the platform literature till date. Our conceptualization of a platform is strongly formatted by the arch-typical examples of iOS/ Apple and the like. The STS corner of information/ knowledge infrastructure has deep affinities with conceptualisations of platforms, notably their generative, open-ended character (Platin et al 2016). Different from platform conceptualisations, however, infrastructure studies perfectly parallel with Henfridsson et al. underscore how digital serve multiple (recombination of) use. As Leigh Star early expressed it, there is never only one infrastructure. For her, this had political connotations. One person's infrastructure is the next one's brick wall (cf Star 1990).

*The issue*: There are two concerns that emerge by drawing on non-IS debates.

First, the paper suggests

> "We refer to paths channeling as the activity of steering value connections, and ultimately value paths, through one particular, or a combination of, resource/s to provide the potential for capturing value as it becomes more center-stage in a particular value space." (p. 14)

The problem with this, drawing on STS, is the strong sense of intentionality assumed by 'steering' value paths. Several have critiqued this. Latour (1999) underscores the inherent 'slight surprise' of action, the inherent side-effects that threaten the intentional main effects (cf. also work on reflexive modernization, Beck et al. 1994).

Second, the paper outlines particular constructs that lend themselves to metrics for measuring design and use recombination (value intensity, scope and overlap). STS contains numerous cautionary tales about the perils of undue confidence in metrics. Our present trust in quantified measures is historically contingent. Quantification has but slowly and with setbacks replaced the qualitative. Crosby (1997) describes how temperature, historically a qualitative phenomenon of 'hot' and 'uncomfortable', gradually got replaced by quantitative measures into degrees centigrade and Fahrenheit. He also describes how quantification in one area (temperature) led to inflated expectations about what could reasonably be quantified when grappling with the quantification of grace and virtue. The relevance, then, to the paper is whether Henfridsson et al. moves too far in the direction of monetizing the notion of value.

Stark (2011) challenges prevailing conceptualizations of value and worth. More specifically, in situations, very much including innovation processes, where you do not know what you are looking for, how do you ascribe value? Rather than a uniform way of evaluating value in the way quantified metrics assume, Stark (ibid., p. 5) argues for multiple, competing and contradictory principles of evaluating value because "there is a principled disagreement about…what is valuable, what is worthy, what counts".

In conclusion, the paper lays out a productive and challenging research programme likely to invigorate the debate on digital innovation in IS research. Formative ideas in the paper resonate with non-IS debates. IS research thrives with the exchange with other disciplines and should be understood as a resource when pursuing the outlined research programme of Henfridsson et al.'s paper.